\documentclass[runningheads]{llncs}
\usepackage{graphicx}
\usepackage{color, xcolor}
\usepackage{subfigure}
\usepackage{caption}
\usepackage{amsmath,amssymb}
\usepackage{booktabs}
\usepackage{multirow}
\usepackage{array}
\usepackage[misc]{ifsym}
\usepackage{threeparttable}
\usepackage[normalem]{ulem}
\usepackage{url}
\usepackage[colorlinks,
            linkcolor=black,       
            anchorcolor=black,  
            citecolor=black,        
            ]{hyperref}

\newcommand{\eqaref}[1]{Eq.(\ref{#1})}
\newcommand{\figref}[1]{Fig.\ref{#1}}
\newcommand{\tabref}[1]{Table~\ref{#1}}
\newcommand{\secref}[1]{Section~\ref{#1}}

\begin{document}
%
\title{DegUIL: Degree-aware Graph Neural Networks for Long-tailed User Identity Linkage}
\titlerunning{DegUIL: Degree-aware User Identity Linkage}
%
%

\author{Meixiu Long \and
Siyuan Chen \and Xin Du \and
Jiahai Wang${(\textrm{\Letter})}$}
\authorrunning{M. Long et al.}

\institute{School of Computer Science and Engineering, Sun Yat-sen University, \\ Guangzhou, China \\
\email{\{longmx7, chensy47, duxin23\}@mail2.sysu.edu.cn, wangjiah@mail.sysu.edu.cn}}
\maketitle              

\begin{abstract}
User identity linkage (UIL), matching accounts of a person on different social networks, is a fundamental task in cross-network data mining. Recent works have achieved promising results by exploiting graph neural networks~(GNNs) to capture network structure. However, they rarely analyze the realistic node-level bottlenecks that hinder UIL's performance. First, node degrees in a graph vary widely and are long-tailed. A significant fraction of \textit{tail nodes} with small degrees are underrepresented due to limited structural information, degrading linkage performance seriously. The second bottleneck usually overlooked is \textit{super head nodes}. It is commonly accepted that head nodes perform well. However, we find that some of them with super high degrees also have difficulty aligning counterparts, due to noise introduced by the randomness of following friends in real-world social graphs.
In pursuit of learning ideal representations for these two groups of nodes, this paper proposes a degree-aware model named DegUIL to narrow the degree gap. To this end, our model complements missing neighborhoods for tail nodes and discards redundant structural information for super head nodes in embeddings respectively. Specifically, the neighboring bias is predicted and corrected locally by two modules, which are trained using the knowledge from structurally adequate head nodes.
As a result, ideal neighborhoods are obtained for meaningful aggregation in GNNs. Extensive experiments demonstrate the superiority of our model. 
Our data and code can be found at \url{https://github.com/Longmeix/DegUIL}.

\end{abstract}


\section{Introduction}
To enjoy diverse types of services, people tend to join multiple social media sites at the same time. 
Generally, the identities of a person on various social platforms have underlying connections, which triggers research interest in user identity linkage (UIL). This task aims to link identities belonging to the same natural person across distinct social networks. 
As an information fusion task, UIL has enormous practical value in many network data fusion and mining applications, such as cross-platform recommendation~\cite{hu2018conet,DBLP:conf/dasfaa/LinCW22}, etc.

\begin{figure}[t]
\centering
    \subfigure[Long-tailed node distribution]{
        \includegraphics[width=0.4\textwidth]{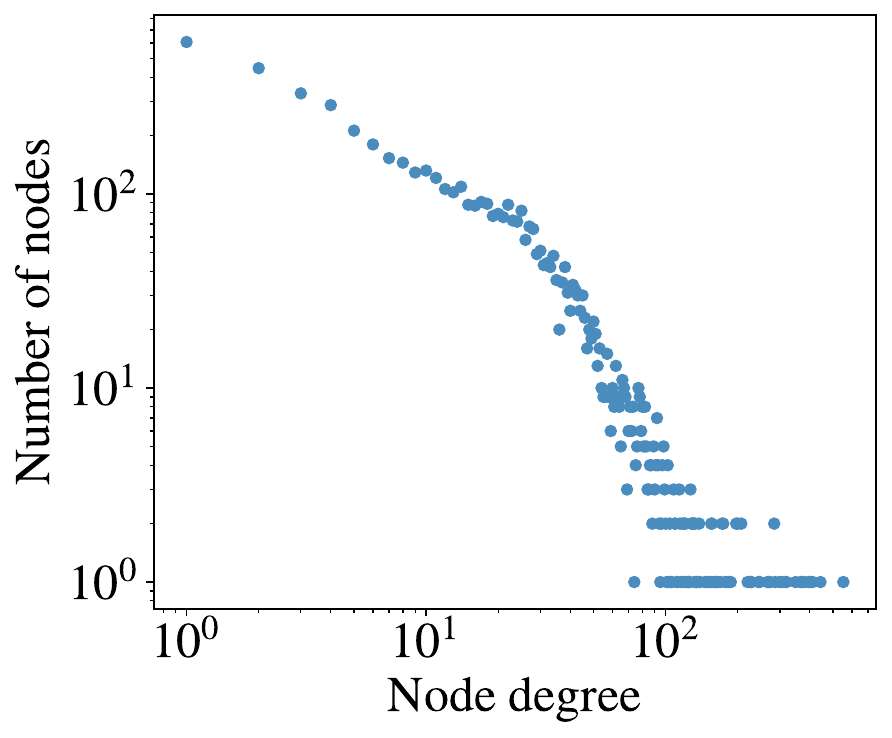}
        \label{fig:power_law}
    }
    \subfigure[MRR w.r.t degrees of test nodes]{
        \includegraphics[width=0.4\textwidth]{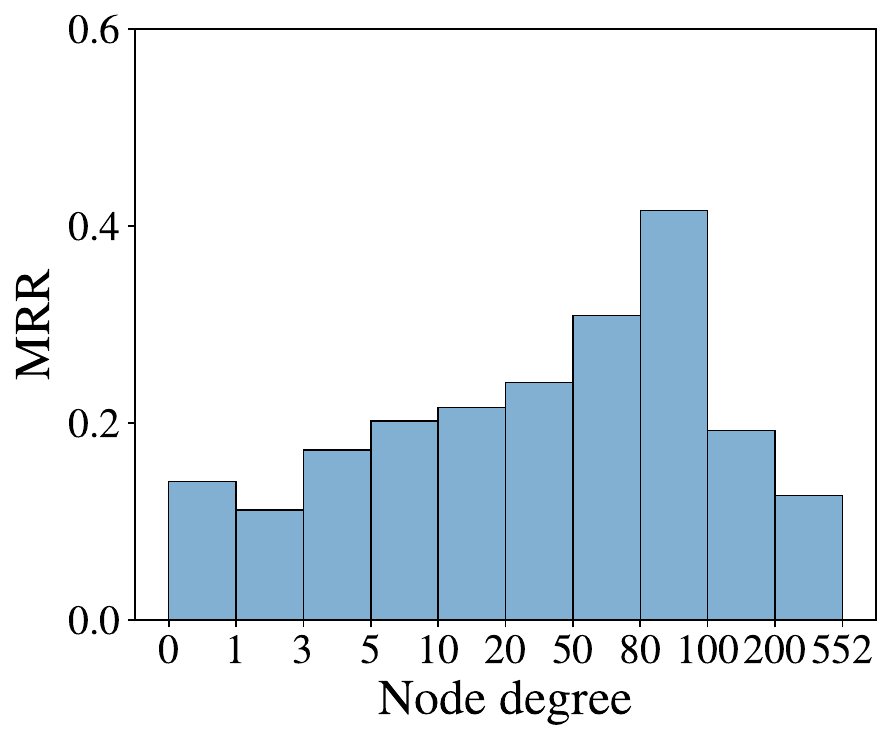}
        \label{fig:low_super_degree}
        
    }
    \caption{ A motivation example on the Foursquare-Twitter dataset with PALE~\cite{DBLP:conf/ijcai/ManSLJC16}. (a)~illustrates the node degree distribution of the Foursquare network, with a large proportion of nodes below 10 degrees. (b)~presents PALE's performance by the degrees of test nodes when 50\% anchors are used for training. Low-degree nodes~$(0, 5]$ and super high-degree nodes~$(200, 522]$ perform worse than the others, indicating these two groups of nodes are the major bottleneck of UIL.
    }
    \label{Fig1_pw}
    
\end{figure}

To date, a corpus of literature has emerged to tackle the UIL problem. Earlier approaches~\cite{DBLP:conf/airs/ZhangKLM14,DBLP:conf/kdd/MuZLXWZ16} aligned users by comparing account profiles such as usernames or post contents. However, such auxiliary information is becoming less accessible and inconsistent due to increased privacy concerns. With the advent of graph neural networks~(GNNs), research attention related to this problem has been shifted to network-structured data.
Although structure-based methods~\cite{DBLP:conf/aaai/TanGCQBC14,DBLP:conf/ijcai/LiuCLL16,DBLP:conf/kdd/ChenYS0GM20} have achieved substantial progress, they rarely doubt whether social networks provide reliable and adequate information for each node.

\paragraph{Realistic Problems.}
In reality, however, social networks are always full of noise and provide scarce structural information, especially in cold-start scenarios with lots of new users.
There are three problems that cannot be ignored.

(1) \textbf{An inherent structural gap exists among nodes.} The number of neighbors varies from user to user in many social networks, and approximately follows a long-tailed distribution, as shown in \figref{fig:power_law}. 
However, existing approaches apply the same learning strategy to all nodes despite their diverse degrees, which hinders the overall linkage performance.
(2) \textbf{The limited neighborhoods of tail nodes hinder the linkage performance.} The performance of structure-aware UIL methods heavily depends on the observed neighborhood. Unfortunately, a significant fraction of low-degree nodes, known as \emph{tail nodes}, connect to few neighbors. In the absence of sufficient structural information, the embeddings of these tail nodes may be unsatisfactory or biased, resulting in inferior performance, as demonstrated in
\figref{fig:low_super_degree}. 
(3) \textbf{Noise hidden in super head nodes exacerbates the quality of representation.}
According to the first-order proximity~\cite{DBLP:conf/www/TangQWZYM15}, UIL works typically assume that friends have similar interests. However, the random nature of users’ behavior in following friends is unavoidable~\cite{DBLP:conf/dmbd/0030W00L022}. Due to this, fraudulent or meaningless edges are hidden in a graph unnoticeably, especially in users with thousands of friends, which is called \emph{super head nodes} in this paper. Small noises in structure can be easily propagated to the entire graph, thereby affecting the embeddings of many others.

All of these realistic issues motivate us to formulate a novel setting for user identity linkage, aimed at improving the linkage performance of tail nodes, which are the most vulnerable and dominant group. In other words, this paper investigates the following research problem: \textbf{how can we effectively link identities for socially-inactive users in a noisy graph?}

\paragraph{Challenges and Our Approach.}
To obtain more competitive embeddings for tail nodes, we need to address three core issues, i.e. data gap, the absence of neighboring information, and noise-filled graphs, which present three challenges.

First, addressing absent neighborhoods poses a dilemma: \emph{tail nodes have no additional information but few neighbors.} This is especially severe if only network structures are available, without accessing additional side information such as profiles or posts on a platform.
Secondly, to defend against the noise in networks, an intuitive idea is to delete fake edges or reduce their negative impacts. However, \emph{how can noise be eliminated while preserving the intrinsic graph structure?} Social networks are full of complicated relationships, making it difficult to discern which edges should be discarded. The above two issues lead to the third challenge: \emph{each node owns both a unique locality and a generality}, which means that bias should be locally corrected without losing the common knowledge across nodes. 

To address these challenges, this paper proposes a degree-aware user identity linkage method named DegUIL to improve the matching of tail identities that account for the majority. 
More concretely, to address the first and second challenges, we utilize the ideal neighborhood knowledge of head nodes to train two modules. They complement potential local contexts for tail nodes and remove redundant neighborhoods of super head nodes in embeddings. Due to this, degree bias is mitigated and their observed neighborhoods are corrected for meaningful aggregation in each GNN layer, thereby improving the quality of node embeddings. 
For the third challenge, two shared vectors are employed across the graph, which adapt to the local context of each node without losing generality. 

    
\paragraph{Contributions.} To summarize, our main contributions are three-fold: 

\begin{itemize}
    \item \textbf{Problem}: This paper highlights that the performance bottlenecks of user identity linkage arise not only from tail nodes but also from super head nodes. The observation motivates us to explore the realistic long-tailed UIL. 
    \item \textbf{Algorithm}: A degree-aware model is proposed to tackle the above two issues, in pursuit of learning high-quality node embeddings for tail nodes' alignment. Our DegUIL corrects the neighborhood bias of the two groups of nodes and thus narrows the degree gap without additional attributes. This strategy brings a novel perspective to the long-tailed UIL problem.
    \item \textbf{Evaluations}: Extensive experiments demonstrate that our model is superior and has significant advantages in dealing with complex networks. 
\end{itemize} 

\section{Related Work}
\paragraph{Structure-based UIL Methods.}
Structure-based methods have become increasingly promising in tackling the UIL problem. Most of them are composed of two major phases: feature extraction and identity matching. Recently, graph neural networks have been well extended into the UIL task~\cite{DBLP:conf/kdd/ChenYS0GM20,DBLP:conf/ecai/ChenWDH20,DBLP:journals/tkde/HongLPT22,DBLP:conf/dasfaa/HuWCD21,li2023semi,DBLP:conf/infocom/0002WTZZL19} and have become mainstream, owing to their powerful capabilities in extracting graph data. For instance, dName~\cite{DBLP:conf/infocom/0002WTZZL19} learns
a proximity-preserving model locally by graph convolutional networks. As simple topology information may be insufficient,
MGCN~\cite{DBLP:conf/kdd/ChenYS0GM20} considers convolutions on both local and hypergraph network structures. While many works neglect topological differences such as low-degree nodes, whose small neighborhood impedes the advance of GNN-based approaches. 
Some recent works in entity alignment are devoted to handling the long-tailed issue by supplementing entity names~\cite{DBLP:journals/ijon/WangWLL22,DBLP:conf/sigir/Zeng00TT20}, or by preventing entities with similar degrees from clustering into the same region of embedded space~\cite{DBLP:conf/www/PeiYHZ19}. 

However, we have not seen a method that rectifies structural bias and narrows degree gap for the realistic UIL task. Different from the existing approaches, our model is dedicated to obtaining high-quality tail nodes' embeddings when no additional side information is available.

\paragraph{Other Long-tailed Problems.}
The long-tailed problem has been studied in many fields~\cite{DBLP:conf/acl/AroraLMKSS18,DBLP:conf/sigir/ChenXLYSD20}, but most of the findings cannot be directly applied to the UIL problem due to differences in problem settings. Two closely related works are Tail-GNN~\cite{DBLP:conf/kdd/LiuN021} and meta-tail2vec~\cite{DBLP:conf/cikm/LiuZFZH20}, which refine feature vectors of tail nodes by transferring the prior knowledge gained from ideal head nodes, leading to a significant improvement in node classification performance. Nevertheless, we observe that not all head nodes are surrounded by ideal neighborhoods in social networks. Structural noise exists in some of very high-degree nodes and impairs performance, as seen in~\figref{fig:low_super_degree}. Therefore, our paper mitigates the noise issue of super head nodes to improve the linkage performance of tail nodes.



\section{Preliminaries}
\subsection{Problem Formulation}
This paper regards a social network as an undirected graph $\mathcal{G}=\left( \mathcal{V},\mathcal{E} \right) $, where $\mathcal{V}=\left\{ v_1,v_2,\dots,v_N \right\}$ is the set of vertices (user identities), $\mathcal{E}=\left\{ e_{ij}=\left( v_i,v_j \right) \right\} \subseteq \mathcal{V}\times \mathcal{V}$ represents the edge set (social connections between users). Each edge $e_{ij}$ is associated with a weight $a_{ij} \in \mathbb{R}$, and $a_{ij}>0$ denotes that node $v_i$ and $v_j$ are connected, otherwise $a_{ij}=0$. Here $\mathbf{A}=\left [a_{ij} \right] \in \mathbb{R}^{N\times N}$ is a symmetric adjacency matrix. $\mathbf{X}\in \mathbb{R}^{N\times d}$ is a feature matrix with $\mathbf{x}_i$ representing the $d$-dimensional feature vector for node $v_i$.
Now our problems are formally defined as below.


\begin{definition}[Super Head Nodes and Tail Nodes]
For a node $v_i \in \mathcal{V}$, let $\mathcal{N}_i$ denote the set of first-order neighbors (neighborhood), and its size $\left| \mathcal{N}_i \right|$ is the degree of $v_i$. Tail nodes have a small degree not exceeding some threshold $D$, i.e. $\mathcal{V}_{\text {tail }}=\left\{v_i:\left|\mathcal{N}_{i}\right| \leq D\right\}$. Nodes with a degree greater than $M$ are super head nodes as $\mathcal{V}_{\text {super}}=\left\{v_i:\left|\mathcal{N}_{i}\right| > M\right\}$. The remaining nodes are called head nodes, i.e. $\mathcal{V}_{\text {head }}=\left\{v_i: D < \left|\mathcal{N}_{i}\right| \leq M \right\}$.
Apparently, $\mathcal{V}_{\text {tail }} \cap \mathcal{V}_{\text {super }} \cap \mathcal{V}_{\text {head }} = \emptyset$. 
\end{definition}

\begin{definition}[User Identity Linkage Aimed at Tail Nodes]
Given two social networks $\mathcal{G}^1$, $\mathcal{G}^2$, and a collection of observed anchor links as inputs, our goal is to identify the unobserved corresponding anchors of tail nodes. Ideally, the matched node should be ranked as top as possible in predicted top-$k$ candidates.
\end{definition}


\subsection{Graph Neural Networks}
A graph neural network with multiple layers transforms the raw node features to another Euclidean space as output. Under the message-passing mechanism, the initial features of any two nodes can affect each other even if they are far away, along with the network going deeper. The input features to the $l$-th layer can be represented by a set of vectors $\mathbf{H}^l=\left\{ \mathbf{h}_{1}^{l},...,\mathbf{h}_{N}^{l} \right\}$, where $\mathbf{h}_{i}^{l}\in \mathbb{R} ^{d_l}$ is $v_i$'s representation in the $l$-th layer. Particularly, $\mathbf{H}^{0} = \mathbf{X}$ is in the input layer. The output node features of the ($l$+1)-th layer are generated as:
\begin{equation}
    \label{eq:GNNs}
    \mathbf{h}_{i}^{l+1}=\mathrm{Agg} \left( \mathbf{h}_{i}^{l}, \left\{ \mathbf{h}_{k}^{l}: k \in
    {\mathcal{N}}_i \right\} ;\theta ^{l+1} \right) 
\end{equation}
where $\mathrm{Agg}\left( \cdot \right)$ parameterized by $\theta^{l+1}$, denotes an aggregation function such
as mean-pooling, generating new node features from the previous one and messages from first-order neighbors. Most GNNs~\cite{DBLP:conf/iclr/KipfW17,DBLP:conf/iclr/VelickovicCCRLB18} follow the above definition.

\section{The Proposed Framework: DegUIL}
\begin{figure*}[t]
    \centering
    \includegraphics[width=\textwidth]{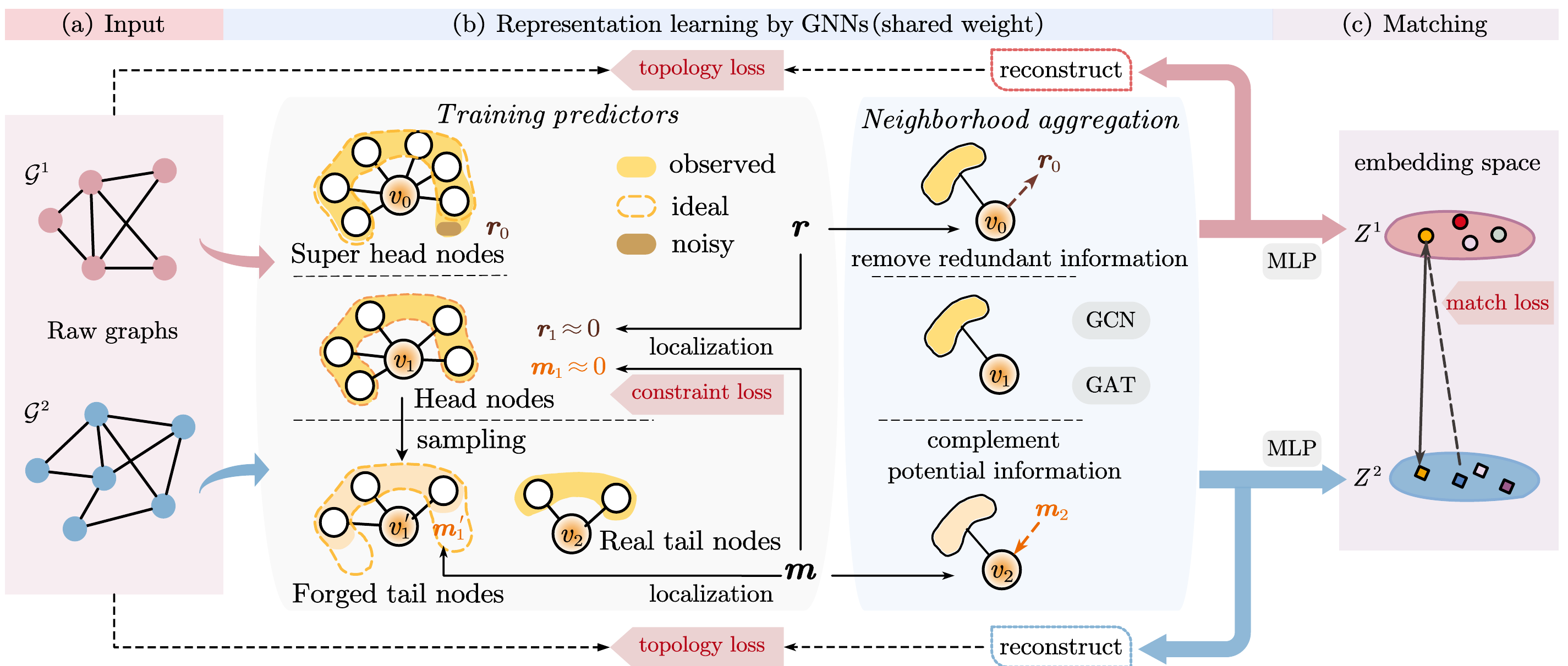}
    \caption{Overview of DegUIL. (a) Inputting two networks; (b) Complementing potential information $m_2$ for tail nodes and removing redundant data $r_0$ for super head nodes to correct their observed neighborhood to be ideal, which improves their representations during aggregation; (c) Mapping two embeddings into a unified space and then matching identities.}
    \label{fig:DegUIL_framework}
\end{figure*}

DegUIL aims to learn high-quality embeddings for tail nodes and super head nodes as a way to enhance linkage performance. Its overall framework is illustrated in~\figref{fig:DegUIL_framework}. As shown in~\figref{fig:DegUIL_framework}(b), we train two predictors named \emph{absent neighborhood predictor} and \emph{noisy neighborhood remover} to predict the neighborhood bias of these two groups of nodes~(\secref{subsection:absent_predictor}--\ref{subsection:redandency_predictor}). 
As a result, tail nodes are enriched by complementing potential neighboring data, and super head nodes are refined by removing noise adaptively, thereby supporting meaningful aggregation~(\secref{subsection:adapt_aggregation}).
Finally, predictors and weight-sharing GNNs are jointly optimized by the task loss and several auxiliary constraints~(Section~\ref{subsection:losses}), for matching identities effectively in~\figref{fig:DegUIL_framework}(c). The target node with the highest similarity to a source anchor node is returned as its alignment result.

\subsection{Uncovering Absent Neighborhood} \label{subsection:absent_predictor}
Neighboring relations connected with tail nodes are relatively few, resulting in biased representations and further hindering linkage results. To solve this problem, we propose an \emph{absent neighborhood predictor} to predict the missing information in their structure, which facilitates subsequent aggregation in each GNN layer. It is trained by exploiting the structurally rich prior learned from head nodes. This component enriches the structural information of tail nodes to obtain better representations as ideal as head nodes.


\subsubsection{Absent Neighborhood Information for Tail Nodes.}
Tail nodes lack structural data owing to a variety of reasons, such as being new users on a social platform. Relationships in networks change dynamically, in other words, tail users may interact with other users in the near future, which can be considered as potential relations. Thus, predicting and completing the latent structural information for tail nodes is reasonable.

More concretely, for a tail node $v_i \in \mathcal{V}_\text{tail}$, the absent information $\mathbf{m}_i$ measures the gap of feature vectors between its observed neighborhood $\mathcal{N}_{i}$ and \emph{ideal neighborhood ${\mathcal{N}}_{i}^*$}, that is,
\begin{equation}
\label{eq:uncover_absent}
\mathbf{m}_i=\mathbf{h}_{ \mathcal{N}_{i}^{*}}-\mathbf{h}_{ \mathcal{N}_{i}}.
\end{equation}

The ideal representation $\mathbf{h}_{ \mathcal{N}_{i}^{*}}$ theoretically contains not only the observed aggregated information from local neighborhoods but also friends that would have been associated with~$v_i$. To construct $\mathbf{h}_{ \mathcal{N}_{i}^{*}}$, we train an absent neighborhood predictor $f_m$ to uncover the missing features caused by limited local contexts. That is, the ideal neighborhood representation of $v_i \in \mathcal{V}_\text{tail}$  can be predicted as $\mathbf{h}_{ \mathcal{N}_{i}^{*}} = \mathbf{h}_{ \mathcal{N}_{i}} +\mathbf{m}_i$. Empirically $\mathbf{h}_{\mathcal{N}_i}$ is represented by a mean-pooling over all nodes in the observed neighborhood, i.e., $\mathbf{h}_{\mathcal{N}_i} = \text{MEAN}(\{\mathbf{h}_{k}: v_k \in \mathcal{N}_i\})$. Now the problem turns into modeling the potential information in a neighborhood.

\subsubsection{Training Absent Neighborhood Predictor.}
The prediction model is learned using the local contexts of head nodes.
Let $\mathbf{m}_{i}^{l}$ be absent neighboring information of node $v_i$ in the $l$-th GNN layer. For a head node $v_j$, its observed neighborhood is regarded as complete and ideal, thus no missing information on its neighborhood. In other words, the representation of $v_j$'s ideal neighborhood can be approximated by $\mathbf{h}_{\mathcal{N}_j}^{l}$, the representation of observed neighborhood $\mathcal{N}_j$ in the same layer. Therefore, we train a prediction model $f_m$ by predicting missing neighborhood information of $v_j$ closed to zero as expected, i.e.  $\lVert\mathbf{m}_j^{l}\|_2\approx 0$. It will
be an auxiliary loss term further discussed in~\secref{subsection:losses}.

However, the training scheme has a major flaw: the abundance of head nodes in training differs from tail nodes in testing. To tackle this problem, \textit{forged tail nodes} are supplemented via edge dropout on head nodes. On each head node, neighbors~($|{\mathcal{N}_i}| \leq D$) are randomly sampled to mimic the real tail nodes. For example, in \figref{fig:DegUIL_framework}(b), $v_{1}^{\prime}$ is a forged tail node generated from the head node $v_1$.

Toward ideal tail nodes representations, a key idea is to uncover the latent information $\mathbf{m}^l_i$ on tail nodes (forged or real), which will be predicted adaptively in~\secref{subsection:adapt_aggregation} to 
correct their observed neighborhoods that may be biased.



\subsection{Removing noisy neighborhood} \label{subsection:redandency_predictor}
As the first step of UIL, learning effective representations for users is crucial. In contrast to tail nodes, super head nodes are structurally rich and even have redundant edges connecting them, since social networks are complex and unreliable. Perturbed neighbors may cause error propagations through the network that drop the final performance~\cite{DBLP:conf/icml/DaiLTHWZS18}. To defend against the damage for further enhancing tail node alignment, we design a \textit{redundant neighborhood remover}.

To be specific, given a super head node $v_i \in \mathcal{V}_\text{super}$, $\mathbf{r}_i$ denotes the embedding redundancy between its observed neighborhood ${\mathcal{N}}_{i}$ and ideal one ${\mathcal{N}}_{i}^{*}$, i.e.,
\begin{equation}
\label{eq:redundancy}
\mathbf{r}_i=\mathbf{h}_{ \mathcal{N}_{i}}-\mathbf{h}_{ \mathcal{N}_{i}^{*}}.
\end{equation}
Our module removes the neighboring bias $\mathbf{r}_i^{l}$ in each layer $l$ to mitigate the error cascade in message aggregation of GNNs. As a result, the ideal neighborhood representation of $v_i $ can be obtained by $\mathbf{h}_{ \mathcal{N}_{i}^{*}}^{l} = \mathbf{h}_{ \mathcal{N}_{i}}^{l} - \mathbf{r}_{i}^{l}$. Similar to the first module, the absent neighborhood predictor, we employ a function $f_{r}$ to predict~$\mathbf{r}_{i}^{l}$. 

To refine an ideal graph, a natural strategy is to eliminate adversarial noise. Many works~\cite{DBLP:conf/kdd/Jin0LTWT20,DBLP:conf/wsdm/TangLSYMW20,DBLP:conf/kdd/ZhuZ0019} delete perturbed edges by graph structure learning or graph defense techniques, but such techniques act on a single network rather than cross-network user matching. Besides, mistakenly deleting a useful edge may lead to cascading defects. Instead, we refine node embeddings directly to distill local structure, which eliminates noise without destroying scarce but valuable relations on tail nodes.
We locally predict redundancy in the following section.

\subsection{Adaptive Aggregation} 
\label{subsection:adapt_aggregation}
\subsubsection{Localization.}
The absent or redundant neighborhood information varies across nodes, hence necessitating fine-grained node-wise adaptation. To capture the unique locality of each node while simultaneously preserving generality across the graph, two globally shared vectors $\mathbf{m}$ and $\mathbf{r}$ (per layer) are introduced. 

Formally, for each node $v_i$ in the $l$-th layer of DegUIL, a locality-aware missing vector $\mathbf{m}_i \in \mathbb{R}^{d_l}$ and a redundant vector $\mathbf{r}_i \in \mathbb{R}^{d_l}$ are customized according to its local context. 
Specifically, the local context information is defined as the concatenation of the node representation with its local observed neighborhood representation, i.e. $\mathbf{c}_{i}^{l}=\left[ \mathbf{h}_{i}^{l}, \mathbf{h}_{\mathcal{N}_{i}}^{l} \right]$.  Then, the absent neighborhood predictor model $f_m$ and noisy neighborhood remover $f_r$ output localized structural information $\mathbf{m}_{i}^{l}$ and $\mathbf{r}_{i}^{l}$, respectively. That is,
\begin{align}
    \mathbf{m}_{i}^{l} &=f_{m}\left(\mathbf{c}_{i}^{l}, \mathbf{m}^{l} ; \theta_{m}^{l}\right) = \boldsymbol{\gamma}_{i}^{l}\odot \mathbf{m}^{l}+\boldsymbol{\alpha}_{i}^{l}, \label{eq:localized_m} \\ 
    \mathbf{r}_{i}^{l} &=f_{r}\left(\mathbf{c}_{i}^{l}, \mathbf{r}^{l} ; \theta_{r}^{l}\right) = \boldsymbol{\gamma}_{i}^{l}\odot \mathbf{r}^{l}+\boldsymbol{\beta}_{i}^{l}, \label{eq:localized_r}
\end{align}
where $\theta_{m}^l$ and $\theta_{r}^l$ are the parameters of $f_m$ and $f_r$ in the $l$-th layer. Element-wise scaling~($\odot$) and shifting~($+$) operations are used to implement the personalization function for each node. 
The scaling vector $\boldsymbol{\gamma}_{i}^{l} \in \mathbb{R}^{d_l}$ can be calculated as $\boldsymbol{\gamma}_{i}^{l} = \mathbf{c}_{i}^{l} \mathbf{W}_{\gamma}^{l} $ with a learnable matrix $\mathbf{W}_{\gamma}^{l} \in \mathbb{R}^{2d_l \times d_l}$. Shift vectors $\boldsymbol{\alpha}_{i}^{l}$ and 
$\boldsymbol{\beta}_{i}^{l}$ are trained using two fully connected networks, respectively.

\subsubsection{Neighborhood Aggregation.}
Our discussion now turns to neighborhood aggregation related to super head nodes and tail nodes. The neighborhoods of head nodes are taken as ideal to follow the standard GNNs aggregation in \eqaref{eq:GNNs}.
In contrast, the embedding vectors of tail nodes are underrepresented and those of super head nodes tend to be noisy. 
Thankfully, our DegUIL complements potential neighboring data for the former and removes local noise for the latter. 

The corrected neighborhoods of these two groups of nodes are ideal for key aggregation in GNN-based methods. In the ($l$+1)-th layer, the standard neighborhood aggregation in~\eqaref{eq:GNNs} is adjusted as follows:
\begin{equation}
    \label{eq:neighborhood_aggregation}
    \resizebox{0.9\columnwidth}{!}{$
    \mathbf{h}_{i}^{l+1}=\text{Agg} \left( \mathbf{h}_{i}^{l}, \left\{ \mathbf{h}_{k}^{l}: v_k \in
    \mathcal{N}_i \right\} \cup \left\{ I\left(v_i \in \mathcal{V}_{\text{tail}}\right) \mathbf{m}_{i}^{l} -I\left(v_i \in \mathcal{V}_{\text{super}}\right) \mathbf{r}_{i}^{l} \right\}; \theta ^{l+1} \right),
    $}
\end{equation}
where $I(\cdot)$ is a 0/1 indicator function based on the truth value of its argument.

\subsubsection{Global and Local Aggregation for UIL.} 
This paper employs two different aggregation strategies to maintain global common knowledge and local structure: 
\begin{equation}
   \mathbf{Z} = \left[ \text{Agg}_{\text{GA}}\left( \mathbf{X, A} \right), \text{Agg}_{\text{LA}}\left( \mathbf{X, A} \right)  \right].
\end{equation}
Here, the global structure aggregator $\text{Agg}_{\text{GA}}\left( \cdot \right)$  observes the whole network by graph convolutional networks~(GCN)\cite{DBLP:conf/iclr/KipfW17}. The local structure aggregator $\text{Agg}_{\text{LA}}\left( \cdot \right)$ acquires specific patterns of nodes' 1-hop neighborhood, implemented by graph attention networks~(GAT)\cite{DBLP:conf/iclr/VelickovicCCRLB18}.
Both of them adopt a two-layer architecture in our method, i.e., $\ell=2$. By stacking aggregation layers, larger area patterns are observed. The final representation $\mathbf{Z}$ is obtained by concatenating the outputs of aggregators.
To preserve the consistency of cross-network node pairs in the embedding space, we apply a shared weight GNN architecture for $\mathcal{G}^1$ and $\mathcal{G}^2$.
In other words, GCN and GAT embed nodes from both the source network and target network via shared learnable parameters.

 


 


\subsection{Training Loss }
\label{subsection:losses}
The whole training process is controlled by three objective terms, 1) topology loss; 2) cross-network mapping loss; and 3) prediction constraints of \eqaref{eq:uncover_absent} and \eqaref{eq:redundancy}. They are described as follows.

\paragraph{Topology Loss.} 
Global topology is preserved by minimizing the weighted difference on all edges between the input and reconstructed networks, i.e.,
\begin{equation}
\mathcal{L}_{s}=\sum_{i=1 }^N \sum_{j=1 }^N b_{ij}\left({a}_{ij}-s_{ij}\right)^{2}=\|(\mathbf{A-\mathbf{S}}) \odot \mathbf{B}\|^{2}_{F}.
\end{equation}
Here, $\mathbf{A}$ represents the adjacency matrix. $\mathbf{S}=[s_{ij}]$ is the new connection matrix where each element is $s_{ij}=\text{Sim} (\mathbf{z}_{i}, \mathbf{z}_{j})$. $\text{Sim} (\cdot , \cdot)$ is the similarity function, cosine similarity here. $s_{ij}$ ranges from $-1$ to $1$, a larger value indicates a stronger social connection between $v_i$ and $v_j$. Moreover, the sampling matrix $\mathbf{B}=[b_{ij}]\in \{0,1\}^{N\times N}$ is used to balance the number of connected and unconnected edges. We adopt a simple uniform negative sampling~\cite{DBLP:conf/uai/RendleFGS09} here, while you are able to make advances by replacing it with better sampling strategies~\cite{DBLP:conf/nips/MikolovSCCD13}.

\paragraph{Cross-network Matching Loss.} 
Existing UIL models~\cite{DBLP:conf/ijcai/ManSLJC16} learn desirable mapping functions $f$ to unify the embeddings of different graphs. 
Formally, given a matched pair $(v^{1}_{i}, v^{2}_a)$ from the set of anchor links $U_a$ and their features $(\mathbf{z}^{1}_{i}, \mathbf{z}^{2}_{a})$, $p=5$ unmatched node pairs $(v^{1}_{i}, v^{2}_{b})$ are sampled uniformly as negative identity links with features $(\mathbf{z}^{1}_{i}, \mathbf{z}^{2}_{b})$. After mapping by functions $f_1$ and $f_2$, the embedding vectors from source network $\mathcal{G}^1$ and target network $\mathcal{G}^2$ are projected to a common embedding space, i.e. $o_i=f_1(z^1_i)$, $o_a=f_2(z^2_a)$ and $o_b=f_2(z^2_b)$, respectively. 
Let $t_{ia}=\text{Sim}(o_i,o_a)$, the loss is defined as:
\begin{equation}
  \mathcal{L} _{t} = \sum\limits_{\left( v_{i}^{1},v_{a}^{2} \right) \in U_a} { \left ( 1 - t_{ia}\right)^{2} }  + \sum\limits_{\left( v_{i}^{1},v_{b}^{2} \right) \notin U_a} {( t_{ib}^{2} + t_{ab}^{2}) }. 
\end{equation}
The objective aims to maximize the similarities of anchor links while minimizing the link probabilities of unmatched identities. $f_1 \left( \cdot; \theta_{f_1} \right)$ and $f_2 \left( \cdot; \theta_{f_2} \right)$ are implemented 
by two multi-layer perceptrons~(MLPs) with learnable parameters $\theta_{f}=\left(\theta_{f_1}, \theta_{f_2}\right)$.

\paragraph{Constraints on Predicted Information.}
For tail nodes, DegUIL aims to complement rather than refine its neighborhood. In contrast, the neighborhood of super head nodes is refined but not enriched. The other nodes' local contexts are regarded as ideal without absence or redundancy. Therefore, both predicted missing data for nodes except tail nodes and noisy information for nodes except super head nodes should be close to zero, which can be formulated as:
\begin{equation}
\resizebox{0.7\textwidth}{!}{$
\mathcal{L}_{p}=\sum_{l=1}^{\ell}\left(\sum_{v_i \notin \mathcal{V}_{\text{tail}}}\left\|\boldsymbol{\mathrm{m}}_{i}^{l-1}\right\|_{2}^{2}+\sum_{v_i \notin \mathcal{V}_{\text{super}}}\left\|\boldsymbol{\mathrm{r}}_{i}^{l-1}\right\|_{2}^{2}\right).
$}
\end{equation}

\paragraph{Optimization.}
For $g=2$ social networks~($\mathcal{G}$), the total loss is is a combined loss:
\begin{equation}
\label{eq:overall_loss}
\mathcal{L}=
\mathcal{L}_{t} +
\lambda \sum_{i}^{g} \mathcal{L}_{s}^{\mathcal{G}^{i}}+ \mu \sum_{i}^{g} \mathcal{L}_{p}^{\mathcal{G}^{i}} .
\end{equation}
Hyperparameters $\lambda$ and $\mu$ balance the importance of 
topology and predicted information constraint. 


\label{appsec:time_complexity}
Here we discuss the computational complexity of DegUIL. Let $N_{\text{max}}=\text{max}\left(\left|\mathcal{V}^1\right|, \left|\mathcal{V}^2\right| \right)$ denote the maximum number of nodes of two input graphs. First, we employ node2vec to generate initial features, resulting in $O(N_{\text{max}})$ complexity. Next, our model employs GCN and GAT to learn powerful representations. In each GNN layer $l$, the overhead involves forging tail nodes, the localization, and the aggregation of absent information and redundant information. Forging tail nodes consumes $O(ND)$ time since we sample up to $D$ neighbors on a head node to forge a tail node, where $D$ is the degree threshold of the tail node; Locally predicting $\mathbf{m}_{i}^{l}$ in \eqref{eq:localized_m} and $\mathbf{r}_{i}^{l}$ in \eqref{eq:localized_r} needs $O(N \bar{D} d_{l}^{2})$ complexity, where $d_l$ is the dimension of the $l$-th layer and $\bar{D}$ is the average node degree. 
Aggregating the corrected neighboring information takes $O(N (\bar{D}+1) d_{l} d_{l-1} )$ time. As $d_l,d_{l-1}$ and the number of GNN layers are small constants, when $\bar{D} \ll N_\text{max}$, the complexity of node2vec and our degree-aware GNNs is $O(N_\text{max})$ for the representation learning process. Overall, the time complexity of our proposed DegUIL is $O(N_{\text{max}})$, i.e., it scales linear time with respect to the number of nodes.

\subsection{Characteristics of DegUIL}
DegUIL is characterized by the following features. 
(1) Unlike most UIL methods that apply the same learning approach to all nodes, our method divides nodes into three groups~(tail/head/super head nodes) according to their degrees. DegUIL considers neighborhood differences and adopts different neighboring bias correction strategies for them to narrow the structural gap by a node-wise localization technique. 
(2) DegUIL predicts and complements potential neighboring information of tail nodes directly, which avoids designing an extra neighborhood translation~\cite{DBLP:conf/kdd/LiuN021} or separates the embedding and refinement processes~\cite{DBLP:conf/cikm/LiuZFZH20}. It eliminates noisy topology of super head nodes implicitly, preventing valuable edges from being deleted by mistake like some graph structure learning methods~\cite{DBLP:conf/kdd/Jin0LTWT20,DBLP:conf/wsdm/TangLSYMW20,DBLP:conf/kdd/ZhuZ0019}. 
(3) We use weight-sharing GNNs instead of two separate GNNs to preserve cross-network similarity and reduce training parameters. 
\section{Experiments}
In this section, we aim to answer the following questions via experiments. 
\textbf{Q1}: How effective is our proposed DegUIL compared with baselines?
    \textbf{Q2}: How does each component of DegUIL contribute to the final results?
    \textbf{Q3}: Is our method compatible with previous data partitions?
    \textbf{Q4}: How much performance does our method improve for nodes in each degree interval?
\subsection{Experimental Settings}
\subsubsection{Datasets.} Two benchmark datasets are employed for evaluation, as summarized in \tabref{tab:tab_dataset}. 
\textbf{Foursquare-Twitter}~(FT), widely used real-world data in previous literature~\cite{DBLP:conf/ijcai/LiuCLL16,DBLP:journals/tkde/LiuLCL20}, provides partial anchor nodes for identity linkage.
\textbf{DBLP17-DBLP19}~(DBLP)~\cite{chen2022mauil} includes two co-author networks, in which a node represents an author, and an edge connects two nodes if they are co-authors of at least one paper. Common authors across two networks are used as the ground truth. We define tail links as anchor links with a node degree of 5 or less.

To simulate a user cold-start scenario where a large number of nodes are tail nodes, anchors containing tail nodes are split into the testing set, and the rest anchor links are used in training.

\begin{table}[t]
  \centering
  \small
  \caption{Dataset statistics.}
    \begin{tabular}{ccccc}
    \bottomrule
    \multicolumn{1}{c|}{Networks} & \multicolumn{1}{c|}{\#Nodes} & \multicolumn{1}{c|}{\#Edges} & \multicolumn{1}{c|}{\#Anchor links} & \multicolumn{1}{c}{\#Tail links} \\
    \hline
    \hline
    \multicolumn{1}{c|}{Foursquare} & \multicolumn{1}{c|}{5313} & \multicolumn{1}{c|}{76972} & \multicolumn{1}{c|}{\multirow{2}[2]{*}{1609}} & \multicolumn{1}{c}{\multirow{2}[2]{*}{443}} \\
    \multicolumn{1}{c|}{Twitter} & \multicolumn{1}{c|}{5120} & \multicolumn{1}{c|}{164919} & \multicolumn{1}{c|}{} &  \\
    \hline
    \multicolumn{1}{c|}{DBLP17} & \multicolumn{1}{c|}{9086} & \multicolumn{1}{c|}{51700} & \multicolumn{1}{c|}{\multirow{2}[2]{*}{2832}} & \multicolumn{1}{c}{\multirow{2}[2]{*}{975}} \\
    \multicolumn{1}{c|}{DBLP19} & \multicolumn{1}{c|}{9325} & \multicolumn{1}{c|}{47775} & \multicolumn{1}{c|}{} &  \\
    \toprule
    \end{tabular}%
  \label{tab:tab_dataset}%
\end{table}%

\subsubsection{Baselines.} To evaluate the effectiveness of DegUIL, we compare it with three kinds of embedding-based baselines, including a conventional representation learning method~(node2vec), state-of-the-art UIL methods and a tail node refinement model~(Tail-GNN). The baselines are described as follows.
\begin{itemize}
    \item \textbf{node2vec}~\cite{DBLP:conf/kdd/GroverL16}: It encodes network topology into a low-dimensional space, whose outputs serve as initial input features to our methods.
    \item \textbf{PALE}~\cite{DBLP:conf/ijcai/ManSLJC16}: This method learns embeddings and predicts anchor links by maximizing the log-likelihood of observed edges and latent space matching. 
    \item \textbf{SEA}~\cite{DBLP:conf/www/PeiYHZ19}: It is a semi-supervised entity alignment method that tries to avoid embedding entities with similar degrees closely by an adversarial training.
    \item \textbf{NeXtAlign}~\cite{DBLP:conf/kdd/ZhangTJXG21}: A semi-supervised network alignment method that achieves a balance between alignment consistency and disparity.
    \item \textbf{Tail-GNN}~\cite{DBLP:conf/kdd/LiuN021}: The GNN framework refines embeddings of tail nodes with predicted missing neighborhood information. Tail-GCN is compared here.
\end{itemize}
Note that node2vec and Tail-GNN are not UIL methods, so the matching process and other settings are the same as ours, for the sake of
fair comparison. All codes come from open-access repositories of the original papers.

\subsubsection{Evaluation Metrics.} Following previous works~\cite{DBLP:conf/kdd/MuZLXWZ16,DBLP:conf/www/PeiYHZ19,DBLP:conf/infocom/0002WTZZL19}, we employ two widely used Hits-Precision~(Hits$@k$) and mean reciprocal
rank~(MRR) as evaluation metrics. $Hits@k=\frac{1}{N} \sum_{i=1}^{N} \frac{k-(hit(v_i)-1)}{k} $, $hit(v_i)$ is the rank position of the matched target user in the top-$k$ candidates. MRR denotes the average reciprocal rank of ground truth results. Higher metric values indicate better performance.

\subsubsection{Setup and Parameters.}
For each method, we set the embedding vector dimension $d=256$ on all datasets. The initial node feature of our method is~generated by node2vec~\cite{DBLP:conf/kdd/GroverL16}. 
We set hyperparameter $\lambda=0.2$ in~\eqaref{eq:overall_loss}, $\mu$ to 0.001 and 0.01 for FT and DBLP datasets respectively. The dimension of hidden layers in $\text{Agg}$ is 64. 
Tail nodes' degree is set to be no greater than 5, i.e. $D=5$, consistent with Tail-GNN. Super head nodes are the top 10\% nodes with the highest degree, thus $M$ is set to \{46, 116, 25, 23\} in four networks (Fourquare, Twitter, DBLP17, DBLP19), respectively. The 2-layer MLP network for matching outputs 256-dimensional embeddings, and the dimension of hidden layers is twice the input length. The optimal hyperparameters for each method are either determined by experiments or the suggestions from the original papers.
All experiments are repeated five times to obtain the average Hits$@k$ and MRR scores. 


\subsection{Result}
\subsubsection{Overwiew of Results~(Q1).}
\begin{table}[t]
  \centering
  \caption{Overall performance. Best result appears in bold and the second best model is underlined except for ablation variants.}
  \setlength{\tabcolsep}{1.6mm}{
   \resizebox{\linewidth}{!}{ 
    \begin{tabular}{c|cccc|cccc}
    \bottomrule
      Dataset   & \multicolumn{4}{c|}{Foursquare-Twitter} & \multicolumn{4}{c}{DBLP17-DBLP19}\\
    \hline
    Metric & Hits$@$1   & Hits$@$10  & Hits$@$30 & MRR   & Hits$@$1   & Hits$@$10  & Hits$@$30  & MRR \\
    \hline
    \hline
    node2vec & 5.43  & 15.08  & 25.49  & 10.93  & 33.18  & 55.10  & 66.52  & 44.17 \\
   PALE & 6.00  & 15.77  & 26.48  & 11.51  & 21.28  & 39.78  & 52.04  & 30.94  \\
   SEA & \underline{6.93}  & 15.89  & 23.94  & 11.80  & \textbf{38.62} & \underline{60.13} & \underline{71.01} & \textbf{49.27}  \\
    NeXtAlign & 6.47  & 12.23  & 16.62  & 9.63  & 36.82  & 59.58  & 70.46  & 48.06  \\
    Tail-GNN & 6.70  & \underline{17.67}  & \underline{28.39}  & \underline{12.66}  & 36.36  & 56.58 & 67.21  & 46.44 \\
    DegUIL & \textbf{9.33}  & \textbf{21.70}  & \textbf{32.81}  & \textbf{16.00}  & \underline{37.59}  & \textbf{60.73}  & \textbf{71.51}  & \underline{48.96} \\
    \midrule
     DegUIL$_{w/o\_{AP}}$ & 8.11  & 19.39  & 30.39  & 14.30  & 36.26  & 59.29  & 70.32  & 47.67 \\
    DegUIL$_{w/o\_{NR}}$ & 8.94  & 20.53  & 31.79  & 15.21  & 37.13  & 59.61  & 70.02  & 48.26 \\
    \toprule
    \end{tabular}%
    }}
  \label{tab:overall_perf}%
\end{table}%

Comparison results on two UIL datasets are presented in \tabref{tab:overall_perf}. 
From the results, we have the following observations. 

-- \emph{DegUIL consistently outperforms other baselines.}  On the Foursquare-Twitter dataset, DegUIL achieves a remarkable relative improvement of 16\%-39\% compared to the best baseline, TailGNN. This is empirical evidence that our method is more effective than previous models in boosting linkage accuracy.
An exception is on the DBLP dataset, where SEA obtains the best Hit@1 and MRR, while DegUIL remains a close runner-up ahead of other baselines. We infer that SEA's technique of encoding relations benefits learning node representations. Besides, with the same mapping process, node2vec is inferior to the GNNs-based Tail-GNN. It demonstrates the power of GNNs in capturing neighboring topology, so mitigating the neighborhood bias to further advance GNNs is significant. 


-- \emph{Degree-aware models perform better than traditional methods.} Node2vec and PALE treat all nodes uniformly without considering the structural disparity such as node degree. As a result, node representations learned by the two simple methods are unsatisfactory for linking user identities. This highlights the importance of degree-aware baselines, which achieve more effective results. However, SEA, NeXtAlign, and Tail-GNN are not specially designed for enhancing super head nodes, their performance still falls short compared to our model. 

-- \emph{DegUIL has a greater advantage in complex long-tailed datasets.} 
Under all evaluation metrics, methods perform worse on the FT dataset than that on the DBLP dataset, despite the former having more known anchor links. One explanation for this discrepancy may be the greater complexity of edge relationships in FT, which makes it challenging to link users in social networks with disparate node degrees. Our model can effectively handle this complex situation, giving it a distinct advantage. Further discussions are in the ablation study.



\subsubsection{Ablation Study~(Q2).} DegUIL comprises two components: an absent neighborhood predictor~(AP) and a noisy neighborhood remover~(NR). To evaluate the contribution of each component, we designed two variants of our model.
\textbf{$\text{DegUIL}_{w/o\_AP}$} does not complement the predicted potential neighborhood for learning tail nodes' embeddings. Another variant model \textbf{$\text{DegUIL}_{w/o\_{NR}}$} does not eliminate the noise from the local structure of super head nodes.  

The results of the ablation study are presented in~\tabref{tab:overall_perf}, which reveals several conclusions. 
First, without AP predicting and complementing absent neighborhoods for tail nodes, UIL performance declines by 1.70\% and 1.29\% in terms of MRR on the FT and DBLP datasets, respectively. This indicates that the limited local context of tail nodes hinders user alignment, and our AP component is proposed as a solution for improving tail node embeddings. 
Second, removing structural noise in super head nodes also contributes to performance. It supports our theoretical motivation that super head nodes are also a challenging group of nodes. 
Notably, the gain of AP is more significant than that of NC on both datasets, suggesting that correcting the neighborhoods of tail nodes offers more substantial alignment benefits. One explanation for this phenomenon is the greater number of tail nodes, compared to super head nodes, which allows them to exert a more considerable influence on the overall performance. 

\begin{figure}[t]
	\centering
	\begin{minipage}[t]{0.3\linewidth}
		\centering
		\includegraphics[width=1.6in]{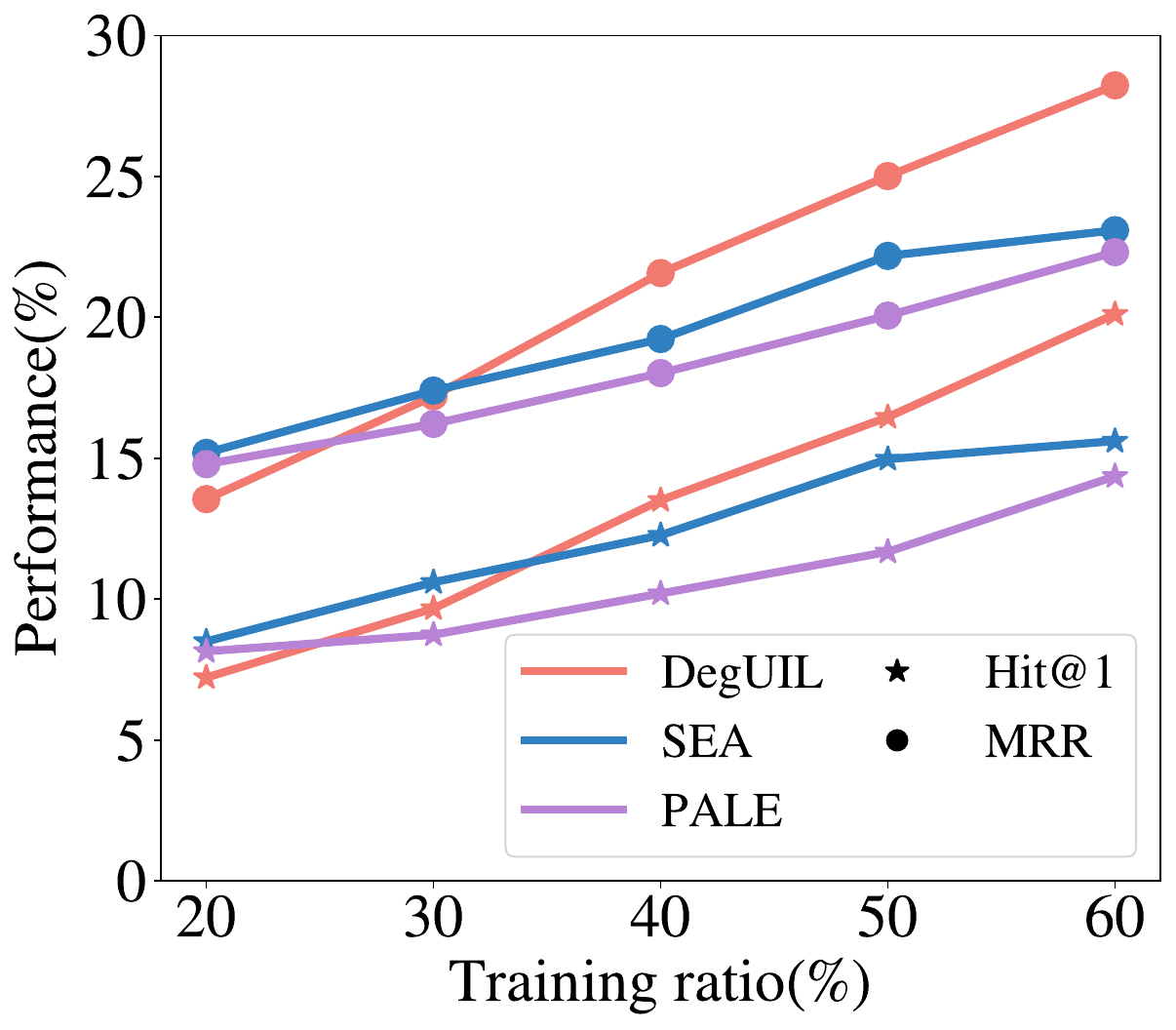}
		\caption{Effect of training ratio on the FT dataset.}
        \label{fig:perf_ratio}
	\end{minipage}
 \hspace{.05in}
	\begin{minipage}[t]{0.67\linewidth}
		\centering
		\includegraphics[width=3in]{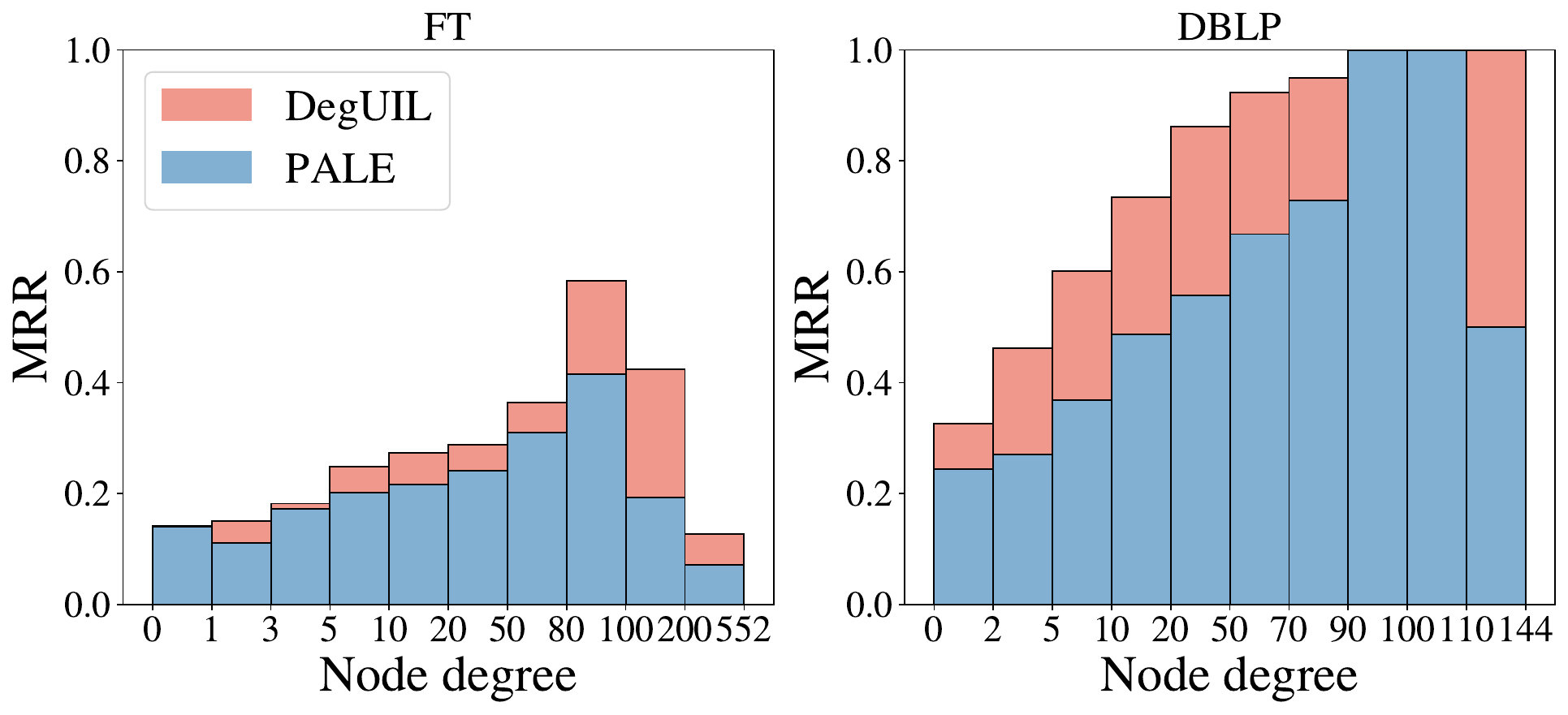}
		\caption{MRR results by degrees.}
    \label{fig:perf_by_degree}
	\end{minipage}
\end{figure}

\subsubsection{Effect on Dataset with Classic Partition~(Q3).}
This paper splits datasets in a novel way to mimic a challenging UIL scenario, i.e. an anchor link without tail nodes is assigned into the training set, otherwise in the testing set. This naturally raises a question: whether DegUIL is compatible with previous ways of data partitioning and still outperforms other baselines under this setting. To answer it, we vary the proportion of labeled anchors for training from 20\% to 60\% with a step of 10\%, and use the rest for testing. Experiments are conducted on the FT dataset with competitive PALE and SEA as comparison methods. 

\figref{fig:perf_ratio} illustrates the Hits@1 and MRR scores. As the training ratio increases, more alignment information is available, enabling all models to discover potential user identities more easily. In most cases, our proposed DegUIL achieves superior performance in both metrics, except when the training data is less than 30\%. This exception arises due to the difficulty of effectively training the GNNs used in DegUIL when labeled supervision is insufficient. In such scenario, SEA and PALE show slight superiority thanks to their semi-supervised way or network extension using observed anchor links. In the future, we will consider semi-supervised or self-supervised training to mitigate the problem of data scarcity.
With more supervision information, DegUIL consistently and significantly outperforms the other two baselines. This means that our degree-aware method is also applicable and competent in the previous data partition.


\subsubsection[Q4]{Evaluation by degree~(Q4).} 
To demonstrate the effectiveness of DegUIL in aligning long-tail entities, we divide the test anchors into multiple groups based on their source node degrees. We compare our method with simple PALE and illustrate their MRR results by degree in~\figref{fig:perf_by_degree}. As hypothesized, low-degree nodes and super high-degree nodes perform worse than those normal nodes with adequate local topology information. This experimental evidence shows that drastic disparities in node degrees could lead to unsatisfactory node representations and biased outcomes. Moreover, DegUIL outperforms PALE across all degree groups in both datasets, validating its effectiveness in handling long-tail issues. While the improvements are smaller on nodes with fewer than two neighbors, given that DegUIL is also constrained by the very 
limited structural information.

\section{Conclusion}
Commonly, node degrees in a social graph are long-tailed, yet UIL works rarely explore the issue of degree bias. We associate the overlooked distribution with UIL performance, observing that the key to improving overall performance is tail nodes and super head nodes.
This paper defines a realistic problem setting and proposes DegUIL to learn high-quality node embeddings by mitigating degree differences in the embedding process through two localized modules.
These modules enrich neighborhood information for tail nodes and refine local contexts of super head nodes. As a result, node representations are improved thanks to the corrected ideal neighborhood.
Extensive experiments show that DegUIL significantly surpasses the baselines. 
In the future, we will consider 
high-order neighborhood and predict structural bias more accurately to enhance our model.

\subsubsection*{Acknowledgement.}
This work is supported
by the National Natural Science Foundation of
China (62072483), and the Guangdong
Basic and Applied Basic Research Foundation
(2022A1515011690, 2021A1515012298).

\bibliographystyle{splncs04}
\bibliography{DegUIL}



\end{document}